\documentclass{ar-1col}
\usepackage[numbers]{natbib}
\usepackage{color}
\jname{Annu. Rev. Nucl. Part. Sci.}
\jvol{69}
\jyear{2019}
\doi{10.1146/((please add article doi))}

\def\lesssim{\mathrel{\hbox{\rlap{\hbox{\lower3pt\hbox{$\sim$}}}\lower-1pt\hbox{$<$}}}}
\def\gtrsim{\mathrel{\hbox{\rlap{\hbox{\lower3pt\hbox{$\sim$}}}\lower-1pt\hbox{$>$}}}}
\newcommand{\alt}{\lesssim}
\newcommand{\agt}{\gtrsim}
\newcommand{\beq}{\begin{equation}}
\newcommand{\eeq}{\end{equation}}
\newcommand{\beqn}{\begin{eqnarray}}
\newcommand{\eeqn}{\end{eqnarray}}

\begin{document}

\markboth{Shibata and Hotokezaka}{Neutron star merger}

\title{Merger and Mass Ejection of Neutron-Star Binaries}

\author{Masaru Shibata,$^1$ and Kenta Hotokezaka$^2$ \affil{$^1$Max
    Planck Institute for Gravitational Physics (Albert Einstein
    Institute), Am M\"ühlenberg 1, PotsdamG-olm, D-14476, German, and
    Center for Gravitational Physics, Yukawa Institute for Theoretical
    Physics, Kyoto University, Kyoto 606-8502, Japan}
    \affil{$^2$Department of 
    Astrophysical Sciences, Princeton University, Princeton, NJ
    08544, USA}}

\begin{abstract}
  Mergers of binary neutron stars and black hole-neutron star binaries
  are one of the most promising sources for the ground-based
  gravitational-wave (GW) detectors and also a high-energy astrophysical
  phenomenon as illustrated by the observations of gravitational waves
  and electromagnetic (EM) waves in the event of GW170817. Mergers of
  these neutron-star binaries are also the most promising site for
  r-process nucleosynthesis. 
  Numerical simulation in full general relativity (numerical relativity) is a unique approach 
  to the theoretical prediction of the merger process,
   GWs emitted, mass ejection process, and
  resulting EM emission. We
  summarize our current understanding for the processes of
  neutron star mergers and subsequent mass ejection based on the
  results of the latest numerical-relativity simulations.  We
  emphasize that the predictions of the numerical-relativity
  simulations agrees broadly with the optical and infrared
  observations of GW170817. 
\end{abstract} 

\maketitle

\tableofcontents

\section{INTRODUCTION}\label{sec1}

Mergers of neutron-star  binaries [binary neutron stars and black
hole-neutron star (BH-NS) binaries] are one of the most promising
sources of gravitational waves (GWs) for ground-based  detectors,
such as Advanced LIGO, Advanced Virgo, and
KAGRA~\cite{Detector1,Detector2,Detector3}. Advanced LIGO
and Advanced Virgo made the first observation of
GWs from a binary neutron star on August 17,
2017 (GW170817)~\cite{LIGO817}.  We expect that these GW
observatories will detect a number of signals from neutron star binaries
in the next few years.

Neutron star mergers are also attracting attention as  promising
nucleosynthesis sites of heavy elements through the rapid neutron
capture process (r-process)~\cite{Lattimer74,Eichler,Thielmann},
because a significant amount of neutron-rich matter is likely to be
ejected during merger (see Refs.~\cite{Davies,
  Ruffert97,Rosswog,Rosswog1} for the pioneering research). In association
with the production of neutron-rich heavy elements in the merger
ejecta, a strong electromagnetic (EM) emission (kilonova/macronova) is
predicted to be powered by the subsequent radioactive decay of the
r-process elements~\cite{Li,Metzger2010,Roberts2011, GBJ2011,
  Oleg,BK2013,TH2013}. This will be an EM counterpart of GWs from neutron star mergers and its detection could be used to
verify the neutron star merger scenario for the origin of r-process elements.
This hypothesis is strengthen by the observation of ultra-violet, optical, and
infrared signals of
GW170817~\cite{LIGO817em,EM0,EM1,EM7,EM8,EM4,EM5,EM6,EM2,EM3, EM111}.  In
addition to kilonovae, a long-lasting synchrotron emission in
multi-wavelengths could arise from the interaction of the merger
ejecta with the interstellar medium (ISM) \cite{NP2011}. 
To detect such
signals is a unique probe to study the velocity profile of the merger
ejecta.  All these facts have encouraged the community of
GW astronomy to theoretically explore the mass
ejection mechanisms, r-process nucleosynthesis, and associated EM
emission in neutron star mergers.

To study  these topics quantitatively, we must clarify the
merger process, subsequent mass ejection, nucleosynthesis and
subsequent decay of heavy elements in the ejecta, and EM emission arising from
the ejecta. 
Numerical-relativity simulations that take into account the
detailed microphysical processes, neutrino radiation transfer, and
magnetohydrodynamics (MHD), are currently our best approach to the problem.
Considerable efforts have been devoted to developing
numerical-relativity simulations for neutron star mergers 
over the past two decades, since the first successful simulation of a
binary neutron star merger in 1999~\cite{S1999,SU2000}. Now, detailed
modeling for the merger phenomena is feasible. In particular, during the
last decade,  researchers have performed
a wide variety of numerical-relativity simulations,
taking into account finite-temperature effects for neutron star
equations of state (EOSs)~\cite{Duez10,Sekig2011}, neutrino
cooling~\cite{Sekig2011,Deaton13,Foucart14,Palenzuela2015} and
neutrino heating~\cite{Sekig15,Foucart2016a}, and MHD
instability~\cite{Kiuchi14,Kiuchi15a,Kiuchi15}, have been performed.
 Numerical relativity has become a robust tool
to study merger phenomena, and it allows us to predict observational
features of neutron star mergers.


The mass ejection processes have been explored with numerical-relativity simulations
since the publications  by Hotokezaka et al.~\cite{Hotoke13a}
for binary neutron stars and by Foucart et al.~\cite{Foucart13} for
BH-NS binaries (see also Bauswein et al.~\cite{Bauswein} for an
approximately general-relativistic work). A variety of
numerical-relativity simulations have been performed to explore the
nature of dynamical
ejecta~\cite{Sekig15,Palenzuela2015,Loverace13,Kyutoku13,Foucart15,Foucart2016,Foucart2016a,Sekig16,Lehner2016,radice2016,Foucart17,Kyutoku17,Dietrich17,Dietrich17b,Bovard,Radice2018}.
These publications have clarified that the mass of the dynamically
ejected matter during merger depends strongly on the EOS, total mass
and mass ratio of the system, and BH spin (for BH-NS
binaries).  For binary neutron stars, the ejecta components have
a somewhat  broad range of electron fraction between
$\approx 0.05$ and $\approx 0.5$ irrespective of the EOS, 
where the electron fraction denoted by $Y_e$ is the electron
number density per baryon number density.
This broad
$Y_e$ distribution is well suited for explaining the abundance
patterns of r-process  elements with mass numbers larger
than $A\sim 90$ observed in the Solar System and  metal-poor
stars~\cite{Wanajo14,radice2016}. By contrast, for BH-NS
binaries, the electron fraction of the dynamical ejecta is always low
($Y_e \alt 0.1$), and hence, heavy r-process elements ($A\gtrsim 130$) are dominantly
synthesized~\cite{Fernandez17}.

After a binary neutron star merger, a BH or massive neutron star (MNS)
surrounded by a dense massive disk (or torus) is formed.  Since 2013,
various simulations for the evolution of such post-merger
remnants have been
performed~\cite{FM13,FM14,Perego,FM15,Just2015,Fernandez17,SM17,Fujiba17,Fetal18}.
These simulations have indicated that a large fraction of mass of
compact disks surrounding the central compact objects is ejected from
the system by a viscous, nuclear recombination, and/or MHD effect.
The mass of this ejecta can be of order $10^{-2}M_\odot$; thus,
it can dominate over the mass of dynamical ejecta, implying that
this ejecta is as important as or even more important than
dynamical ejecta to power EM emission.

The purpose of this article is to review the merger process and mass
ejection mechanisms in neutron star merger, and to summarize possible
EM emission from the merger ejecta.  This review is organized as
follows. In \S~\ref{sec2}, we summarize processes of the merger and
post-merger phases of neutron star binaries based on the latest
results of numerical-relativity simulations. In \S~\ref{sec3}, we
describe mass ejection processes during merger and from the
post-merger remnants. In \S~\ref{sec4}, we list the representative EM
signals (ultra-violet, optical, infrared, and radio signals) emitted from the ejecta
of neutron-star mergers. Finally, in \S~\ref{sec5}, we note that the
optical and infrared signals of GW170817 are consistent broadly with
the prediction by numerical relativity. 


\section{SCENARIOS FOR NEUTRON-STAR MERGER AND POST MERGER}\label{sec2}

\begin{figure}[t]
\vspace{-1mm}
\includegraphics[width=110mm]{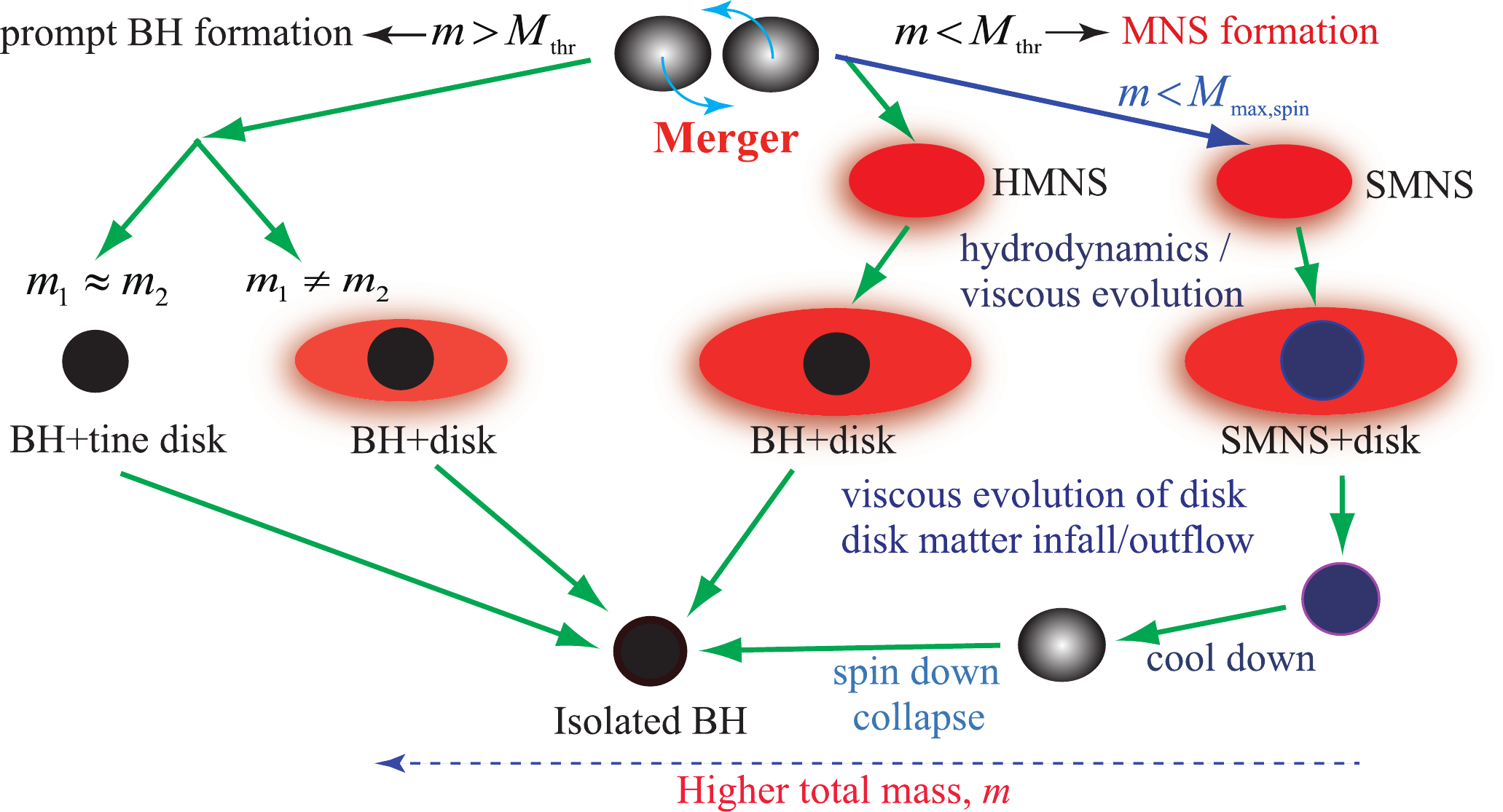}
\caption{A summary for the merger and post-merger evolution of binary
  neutron stars. $M_{\rm thr}$ and $M_{\rm max,spin}$ denote the
  threshold mass for the prompt formation of a BH and the maximum mass
  of rigidly rotating cold neutron stars, respectively. Their values
  are likely to be $M_{\rm thr} \agt 2.8M_{\odot}$ and $M_{\rm
    max,spin}\agt 2.4M_{\odot}$. For the total mass $m > M_{\rm thr}$,
  a BH is formed in the dynamical timescale after the onset of merger,
  and for the nearly equal-mass case, $m_1 \approx m_2$, the mass of
  disks surrounding the BH is tiny $\ll 10^{-2}M_{\odot}$, while it
  could be $\agt 10^{-2}M_{\odot}$ for a highly asymmetric system with
  $m_2/m_1 \alt 0.8$. For $M_{\rm max,spin} < m < M_{\rm thr}$, a {\em
    hypermassive} neutron star (HMNS) is formed, and it subsequently
  evolves through several angular-momentum transport processes,
  leading to eventual collapse to a BH surrounded by a disk (or
  torus). See Refs.~\cite{BSS00,NR2016} for the definition of the HMNS
  (and SMNS referred to below).  When $m$ is close to $M_{\rm thr}$,
  the lifetime of the MNS is relatively short, while for smaller
  values of $m$ toward $M_{\rm max,spin}$, the lifetime is longer. For
  the longer lifetime, the angular-momentum transport process works
  for a longer timescale, and the disk mass could be $\agt
  0.1M_{\odot}$, whereas for a short lifetime, it could be $\sim
  10^{-2}M_{\odot}$ or less.  For $m < M_{\rm max,spin}$, a {\em
    supramassive} neutron star (SMNS) is formed and it will be alive
  for a dissipation timescale of angular momentum which will be much
  longer than the cooling timescale $\sim 10$\,s. Note that MNS
  denotes either a SMNS or a HMNS.
\label{fig1}}
\end{figure}

The fate of neutron star mergers depends on the mass ($m_1, m_2$) and
spin of binary components, and on the neutron star EOS.  For binary
neutron stars, for which the effect of their spin is minor, the total
mass ($m=m_1+m_2$), the mass ratio ($q=m_2/m_1~(\leq 1)$) of the
system, and the EOS are the key quantities for determining the merger
remnant.  For BH-NS binaries, the BH spin as well as the mass ratio
and neutron-star EOS are the key quantities. In the following two
subsections, we classify the remnants formed after neutron-star
mergers.

\subsection{Binary Neutron Stars}\label{sec2.1}

Figure~\ref{fig1} summarizes the possible remnants and their evolution
processes for mergers of binary neutron stars. Broadly speaking, there
are two possible remnants formed immediately after the onset of
merger; BH and MNS. A BH is formed if the total mass $m$ is so high
that the self gravity of the merger remnant cannot be sustained by the
pressure associated primarily with the repulsive force among nucleons
and centrifugal force due to rapid rotation associated with the
orbital angular momentum of the premerger binary.

In the last decade,
simulations were performed employing a variety of neutron-star
EOSs (e.g., \cite{Shibata050,Shibata05,Kiuchi09,Hotoke11,Hotoke13a,Hotoke13b,Hotoke13c,Hotoke13d,ber2015,Sekig15,Palenzuela2015,radice2016,Foucart2016,Foucart2016a,Ciolfi}),
 of which the maximum mass of a non-rotating neutron star is consistent with the existence of  two-solar-mass neutron
stars~\cite{demorest10,Anton13}.
An important finding for these simulations is that for $m \alt
2.8M_\odot$, the remnant is, at least temporarily, an MNS not a BH
irrespective of the EOS employed.

The total mass of nine Galactic binary neutron stars for which
the merger time is less than a Hubble time of $\sim 13.8$\,Gyr is in the
range between $\approx 2.50M_\odot$ and
$2.88M_\odot$~\cite{tauris,lorimer}. Among them, seven objects have a total mass
smaller than $2.75M_\odot$, suggesting that, for the typical total mass
of binary neutron stars, an MNS should be formed after merger (at
least temporarily). In fact, the total mass of the binary neutron star GW170817 is in the middle of the above range,
$2.74^{+0.04}_{-0.01}M_{\odot}$, for a low spin prior~\cite{LIGO817}.

For $m \agt 2.8M_\odot$, a BH could be formed immediately after
merger, although the threshold mass for the prompt BH formation
depends strongly on the EOS.  The dimensionless BH spin, $\chi$, in
the prompt BH formation case is $\approx 0.8$~\cite{Kiuchi09}.
The remnant BH in this formation channel is not surrounded by a
massive disk if the mass ratio, $q$, is close to unity.  The mass of
the disk surrounding the BH increases with the decrease of $q$, and in
the presence of a significant mass asymmetry, $q \alt 0.8$, the disk
mass could be $\agt
10^{-2}M_\odot$~\cite{Shibata05,Kiuchi09,Hotoke13b}.  The disk 
 is evolved by MHD processes, in particular by the effect of MHD
turbulence induced by magnetorotational instability (MRI)~\cite{MRI}
or viscous process (see \S~\ref{sec2.2}). During the MHD or viscous
evolution of the disk, a short gamma-ray burst (sGRB) jet may be launched
from the vicinity of the BH by pair-annihilation processes of
neutrinos emitted from the inner region of the
disk~\cite{RM92,RJ96,DiMatteo02,Lee05,SRJ06,SST07,Chen07} and/or by
the effect of strong magnetic fields such as the Blandford-Znajek
mechanism~\cite{BZ,McKinney06,PRS15,Ruiz16}.  The viscous angular
momentum transport process also drives mass ejection in the viscous
timescale of the disk (see \S~\ref{sec3} for details).

In the case of  MNS formation, the MNS's evolution is determined by several
processes. Soon after its formation, the gravitational torque
associated with nonaxisymmetric structure of the merger remnant plays
an important role for transporting angular momentum from the MNS to
the surrounding matter (e.g., Ref.~\cite{Hotoke13b}). This process
reduces the angular momentum of the MNS. If it is marginally stable
against gravitational collapse, the MNS collapses to a BH due
to this process in $\sim 10$\,ms. The resulting system is a spinning
BH of $\chi \sim 0.6$--0.7 surrounded by a disk of
mass $10^{-2}$--$10^{-1}M_\odot$~(e.g., Refs.~\cite{Sekig16,Dietrich17}).

On a longer timescale, viscous effects resulting from MHD turbulence
are likely to play a key role in the evolution of the
MNS~\cite{Duez04,Shibata17a,Fujiba17}. At its formation, the MNS is
differentially rotating. Furthermore, it should be strongly magnetized
and in an MHD turbulence state exciting a turbulent viscosity, because
 a velocity-shear layer is formed at the contact surfaces of the merging two
neutron stars and the Kelvin-Helmholtz instability
occurs~\cite{PR,Kiuchi14,Kiuchi15a,Kiuchi17}. This instability generates a
number of small-size vortexes  near the shear layer, and consequently, 
magnetic fields are wound up by the vortex motion, which enhances the
magnetic-field strength on a timescale much shorter than the dynamical
timescale of the system, $\sim 0.1$\,ms. Note that the growth timescale
of the Kelvin-Helmholtz instability~\cite{Chandra61} is $\tau_{\rm KH}
\sim 10^{-7}(\lambda/1\,{\rm cm})$\,ms for the wavelength $\lambda$
because the typically velocity at the onset of merger is $\sim
10^{10}$\,cm/s. Because of the presence of the differential rotation
and turbulent viscosity, the angular momentum in the MNS should be
transported outward, and as a result, the MNS is likely to settle to a
rigidly rotating
state~\cite{Duez04,Shibata17a,Fujiba17}. Simultaneously, a massive
disk surrounding the MNS is formed because of the angular momentum
transport.  If this angular momentum transport significantly weakens
centrifugal force in its central region, the MNS could collapse to a
BH. Using the $\alpha$-viscous prescription for the turbulent
viscosity~\cite{SS73}, one can estimate the viscous timescale as
\beq
\tau_{\rm vis,MNS} \approx 20\,{\rm ms}
\left({\alpha_{\rm vis} \over 10^{-2}}\right)^{-1}
\left({c_s \over c/3}\right)^{-1}
\left({R \over 15\,{\rm km}}\right)^{2}
\left({H \over 10\,{\rm km}}\right)^{-1}, \label{tvis0}
\eeq
where $\alpha_{\rm vis}$ is the dimensionless viscous parameter, $c_s$
 is the sound velocity, $c$  is the speed of light, $R$ is the
equatorial radius of the MNS, and $H$ is the  maximum size of
the turbulent vortex. If a turbulent state is sufficiently developed, then
$\alpha_{\rm vis}$ will become of the order $10^{-2}$
according to the latest results of high-resolution MHD simulations for
accretion disks~\cite{alphamodel,suzuki,local}. 

If the lifetime of the MNS is longer than $\tau_{\rm vis,MNS}$
(i.e. MNS mass is not very large), then it will be evolved through the viscous
accretion from the disk and cooling by neutrino
emission~\cite{Fujiba17}. The viscous timescale of the disk is
 written approximately as
\beq
\tau_{\rm vis,disk} \approx 0.5\,{\rm s}
\left({\alpha_{\rm vis} \over 10^{-2}}\right)^{-1}
\left({c_s \over c/10}\right)^{-1}
\left({R_{\rm disk} \over 50\,{\rm km}}\right)
\left({H/R_{\rm disk} \over 1/3}\right)^{-1}, \label{tvis}
\eeq
where $R_{\rm disk}$ is the typical disk radius.
The neutrino cooling timescale for MNSs is 
\beq
\tau_{\nu}\approx {U \over L_{\nu}}
=10\,{\rm s} 
\left({U \over 10^{53}\,{\rm erg}}\right)
\left({L_\nu \over 10^{52}\,{\rm erg/s}}\right)^{-1}, \label{tnu}
\eeq
where $U$ is the thermal energy of the MNS and $L_\nu$ is the total neutrino
luminosity. Note that at the formation of the MNS, $L_\nu \agt
10^{53}$\,erg/s~\cite{Sekig2011,Sekig15,Palenzuela2015,radice2016},
because shock heating at merger significantly increases its
temperature, but in $\sim 100$\,ms after the formation, $L_\nu$ is likely to
decease to $\lesssim 10^{53}$\,erg/s~\cite{Fujiba17}. Thus, if the
viscous accretion onto the MNS or the neutrino cooling has a
significant effect and the MNS is marginally stable against gravitational
collapse, the MNS would collapse to a BH on either of these
timescales.

If the MNS mass is sufficiently low, it will not collapse to a BH in $\sim
10$\,s. In this case, the MNS is likely to settle to a rapidly and
rigidly rotating cold neutron star (a so-called SMNS). The maximum
mass of the SMNS is by $\sim 0.4M_\odot$ increased by the rigid
rotation if it is rotating nearly with the maximum angular velocity,
$\sim \sqrt{GM_{\rm MNS}/R^3}$~\cite{FIP86,CST94}, where $M_{\rm MNS}$
denotes gravitational mass of the MNS and $G$ is the gravitational
constant. For example, if the maximum mass of a cold spherical neutron star is 
$2.2M_\odot$, then the maximum mass of the SMNS would be $\sim 2.6M_\odot$,
so that the self gravity of the SMNS
could be sustained. However, because a SMNS formed in merger is magnetized, its
rotational kinetic energy is subsequently dissipated through the magnetic
dipole radiation if a force-free magnetic field is established outside
the SMNS. Assuming the presence of dipole magnetic radiation with
the luminosity $L_{\rm B}$, the spin-down timescale of the SMNS is
\beqn
\tau_B \approx {T_{\rm rot} \over L_{\rm B}}
\approx 650\,{\rm s}
\left({B_p \over 10^{15}\,{\rm G}}\right)^{-2}
\left({M_{\rm MNS} \over 2.5M_\odot}\right)
\left({R \over 15\,{\rm km}}\right)^{-4}
\left({\Omega \over 7000\,{\rm rad/s}}\right)^{-2}, \label{tb}
\eeqn
where $T_{\rm rot} (\sim 0.3 M_{\rm MNS}R^2\Omega^2)$ is 
rotational kinetic energy, $B_p$ is the magnetic-field strength of the
SMNS pole, and $\Omega$ is the angular velocity of the SMNS. Here, we have assumed
that the magnetic-field strength would be significantly enhanced at
merger.  This estimate shows that the rotational kinetic energy could
be dissipated in $\sim 10^3$\,s. After the dissipation of its 
rotational kinetic energy, the SMNS should collapse to a BH.

\subsection{Black Hole-Neutron Star Binaries}\label{sec2.2}

\begin{figure}[t]
\vspace{-1mm}
\includegraphics[width=110mm]{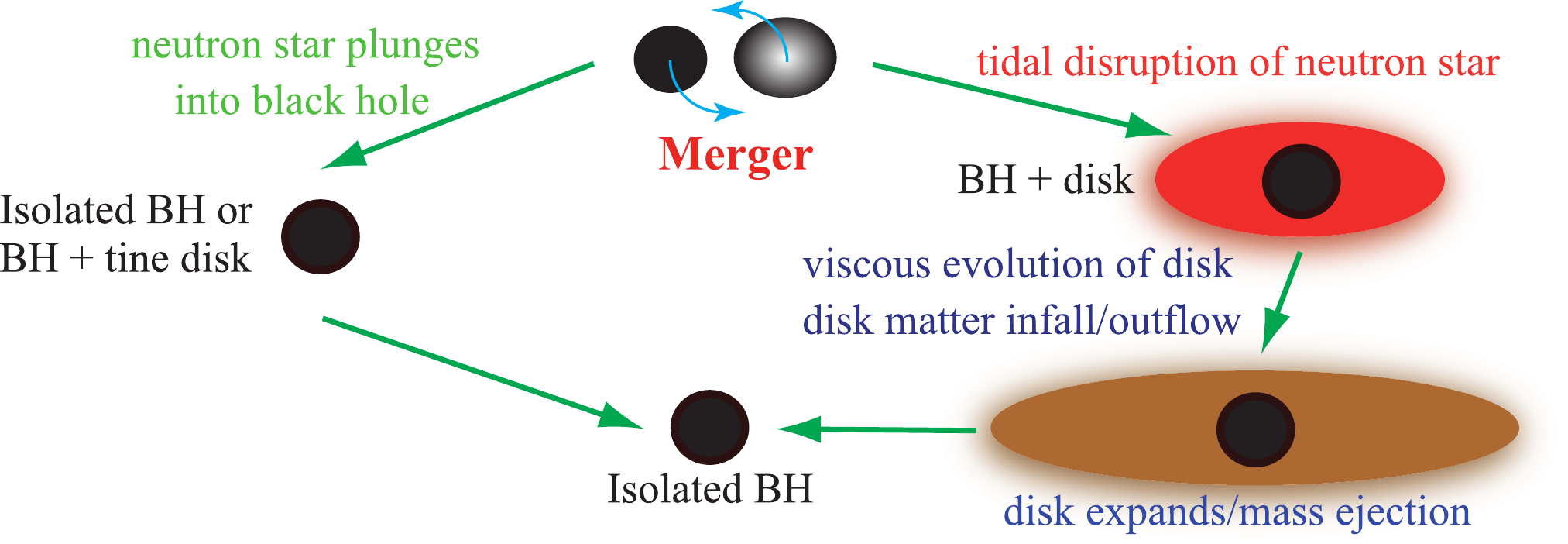}
\caption{A summary for the merger and post-merger evolution of BH-NS
  binaries. This system has two possible fates; the neutron
  star is tidally disrupted or not by the companion BH. For the case
  of tidal disruption, the remnant is a spinning BH surrounded by a
  disk.  The evolution process of the BH-disk system is essentially
  the same as that for binary neutron star mergers. 
\label{fig2}}
\end{figure}

Figure~\ref{fig2} summarizes the possible remnants and their evolution
processes expected for mergers of BH-NS binaries. BH-NS binaries have two
possible fates: Either the neutron star is tidally disrupted before it is
swallowed by the BH or it is  swallowed  by the
BH without disruption~\cite{NR2016}. For the latter case, essentially  disk is not
formed and no matter is ejected, and there is no or weak EM emission.

Tidal disruption of a neutron star occurs if the tidal force by
BHs is stronger than the self-gravity of the neutron star. Assuming
Newtonian gravity, the condition is approximately written as $GM_{\rm
  BH}R_1/r^3 > GM_{\rm NS}/R_1^2$. Therefore,
\beq
\left({GM_{\rm BH} \over c^2r}\right)^{3/2}
\left({M_{\rm NS} \over M_{\rm BH}}\right)
\left({R_1 \over G c^{-2} M_{\rm NS}}\right)^{3/2} > 1, \label{tidaldis}
\eeq
where $r$ is the orbital separation, $M_{\rm BH}$ and $M_{\rm NS}$ are
the mass of the BH and neutron star, and $R_1$ is the semi major axis
of the neutron star. $R_1$ is by a factor of $\sim
1.5$ larger than the neutron-star radius, $R_{\rm NS}$, at the onset
of tidal disruption. For tidal disruption, the condition of
Eq.~\ref{tidaldis} should be satisfied before the neutron-star orbit
reaches the innermost stable circular orbit (ISCO) around the BH, at
which $r=\xi Gc^{-2} M_{\rm BH}$, where $\xi=6$ for non-spinning BHs
and $\xi=1$ for extremely rapidly spinning BHs (which is corotating
with the binary orbit). Here, we have assumed that $Q=M_{\rm BH}/M_{\rm NS}$
is large enough that we can ignore the tidal deformation effect of neutron
stars to the orbital motion. Then, we can rewrite Eq.~\ref{tidaldis} as
\beq
\left({\xi \over 6}\right)^{-3/2}
\left({Q \over 7}\right)^{-1}
\left({R_1 \over 10 Gc^{-2} M_{\rm NS}}\right)^{3/2} > 3.25. \label{tidaldis1}
\eeq
We note that $Gc^{-2}M_{\rm NS} \approx 2.0 (M_{\rm
  NS}/1.35M_\odot)$\,km.  This condition indicates that tidal
disruption occurs for low values of $\xi$ (i.e., for rapidly
spinning BHs) or for low values of $Q$, if the BH 
spin is not very large. Since the value of $Q$ is likely to be higher
than $\approx 4$ for the typical neutron-star mass of 1.3--1.4$M_\odot$,
we find that a high-spin BH is needed for  tidal
disruption of neutron stars.

Numerical-relativity simulations have shown that for the case in which a
neutron star is tidally disrupted, an accretion disk is subsequently
formed around a spinning
BH~\cite{SU06,Kyutoku10,Duez10,Kyutoku11,Foucart11,Foucart12,Deaton13,Foucart13,Loverace13,Kyutoku13,Foucart14,Foucart15,Kawaguchi15,Kyutoku15,Kiuchi15,Foucart17,Kyutoku17}.
Also, a fraction of neutron-rich matter is ejected from the system
(see \S~\ref{sec3.2})~\cite{Foucart13,Kyutoku15,Foucart17,Kyutoku17}.
The disk mass, $M_{\rm disk}$, depends strongly on $Q$, $R_{\rm NS}$,
and BH spin. Among these three parameters, the BH spin is the most
substantial. For example, for a dimensionless BH spin, $\chi =0.75$,
with $R_{\rm NS}\approx 13$\,km, $M_{\rm disk}$ can be $\sim 10\%$ and
$20\%$ of $M_{\rm NS}$ for $Q=7$ and 3, respectively~\cite{Kyutoku15}.
For $\chi=0.9$, $M_{\rm disk}$ is $\sim 20\%$ of $M_{\rm NS}$ for
$Q=7$ and $R_{\rm NS} \approx 13$\,km~\cite{Foucart13}.  Loverace et
al. find that for $\chi=0.97$, with $Q=3$ and $R_{\rm NS} \approx
14$\,km, $M_{\rm disk}$ can be $\sim 0.5M_{\rm
  NS}$~\cite{Loverace13}.

Next, we turn our attention to MHD/viscous evolution of  a disk surrounding
a rapidly spinning BH after a  BH-NS merger.  Such an accretion disk has nearly Keplerian
motion (i.e., differential rotation) and should have magnetic fields
originating in the neutron star's magnetic fields. Thus, the disk
is unstable to the MRI, and as a result, it is likely to be in a
turbulent state~\cite{MRI}, enhancing turbulent
viscosity~\cite{alphamodel,suzuki,local}. Therefore, the BH accretion disk
evolves through the viscous process on the timescale
of Eq.~\ref{tvis}. Specifically, viscous heating and angular
momentum transport, together with neutrino cooling, are the key
processes. Through viscous angular momentum transport,
matter in the inner part of the disk falls into the BH while its outer
part gradually expands along the equatorial plane. Viscous
heating increases the temperature of the disk  to $\sim 1$--10~MeV,
leading to appreciable neutrino
emission~\cite{RJ96,DiMatteo02,Lee05,SRJ06,SST07,Chen07,SM17,Fetal18}. If
the density of the disk is sufficiently high, $\agt 10^{11}\,{\rm
  g/cm^3}$, then the optical depth to neutrinos is large enough to avoid
free-streaming escape, suppressing 
neutrino emissivity. In this phase, the temperature of the disk is determined by
the condition that the timescales of the neutrino cooling and viscous
heating approximately agree  with each other. Throughout the evolution of the
system, the density of the disk decreases because of the mass infall
into the BH together with expansion of the disk by the viscous angular
momentum transport. Then, the optical depth of the disk to neutrinos
decreases~\cite{FM13}. In this later phase,  adiabatic expansion of
the disk (not the neutrino cooling) as well as  infall into the BH
becomes the primary cooling process while  viscous heating is
always the dominant heating process.  This late-phase adiabatic
expansion of the disk eventually drives mass ejection (see
\S~\ref{sec3.3}).

The vicinity of spinning BHs is likely to be the site for high-energy
phenomena for two reasons.  First, the 
temperature of disks near BHs can be quite high $\agt 10$~MeV, and
hence high-energy neutrinos are copiously emitted. Because of the
high temperature, the disks can be geometrically thick,
so an appreciable fraction of neutrinos are emitted toward the
rotational axis of the spinning BHs.  This enhances the pair
annihilation of neutrinos and their anti-neutrinos, leading to 
pair production of electrons and positrons, which could subsequently
produce $\gamma$-rays through pair annihilation. If the total energy of
electrons and positrons is high enough, they could be the engine for
driving an
sGRB~\cite{RM92,RJ96,DiMatteo02,Lee05,SRJ06,SST07,Chen07,SM17}

Second, as  mentioned above, the BH accretion disk is likely to be strongly
magnetized due to the MRI. If the resulting magnetic pressure is high
enough to blow off the matter in the vicinity of the disk, then MHD outflow
could be driven. Subsequently, poloidal magnetic fields are likely to
be formed near the spinning BH and some of the magnetic field lines
would penetrate the BH horizon. In such a magnetic-field
configuration, the rotational kinetic energy of the spinning BH could be
extracted by the Blandford-Znajek mechanism~\cite{BZ}.  If the extracted
energy is well collimated toward the polar direction and leads to
relativistic jets, an sGRB may be produced~\cite{Kiuchi15,McKinney06,PRS15,Ruiz16}.

\section{MASS EJECTION FROM NEUTRON-STAR MERGERS}\label{sec3}

\begin{figure}[t]
\vspace{-15mm}
\includegraphics[width=125mm]{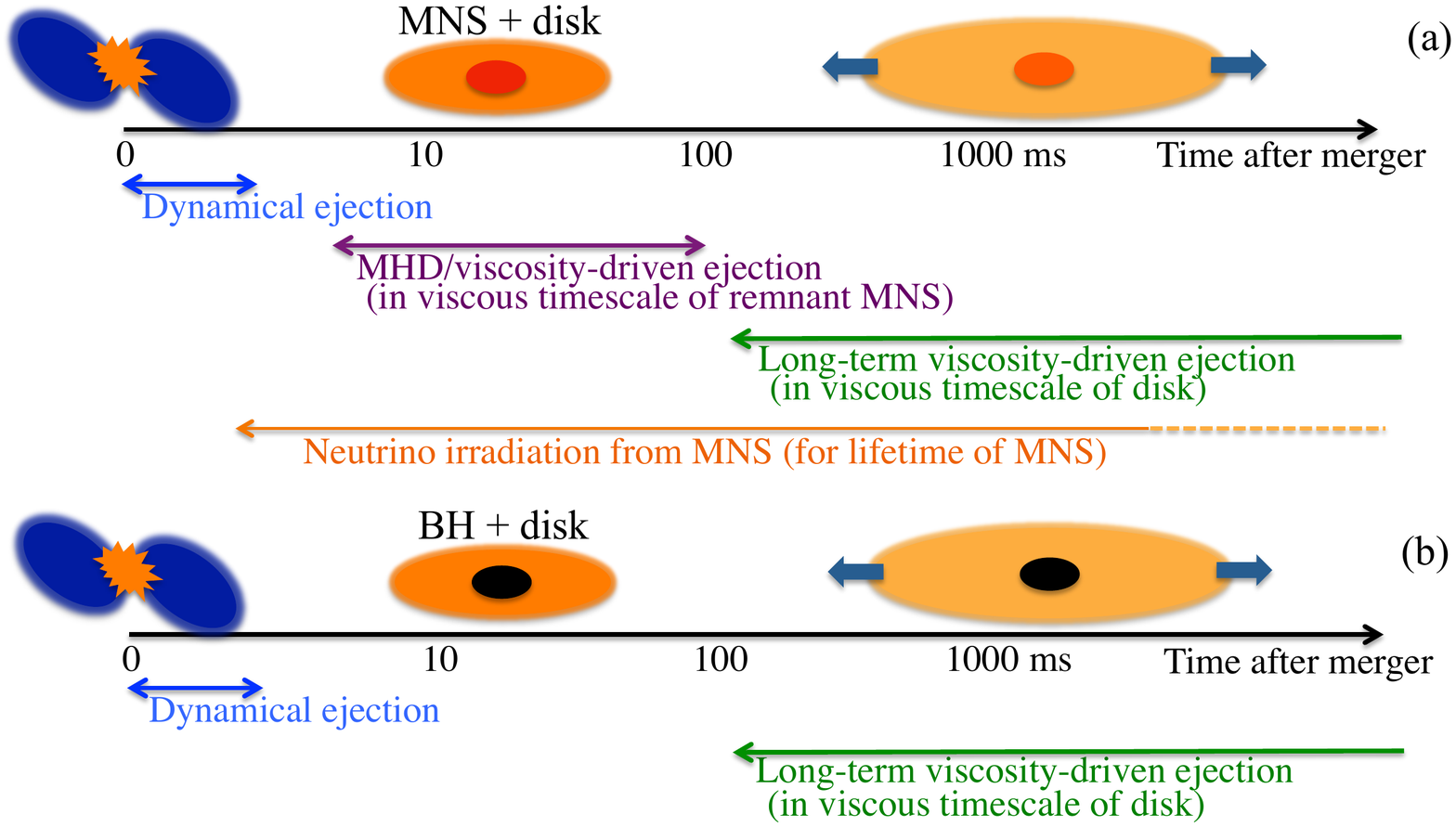}
\vspace{-15mm}
\caption{Mass ejection mechanisms during and after merger of binary
  neutron stars. Soon after the onset of merger, dynamical mass
  ejection occurs in the timescale of ($\alt 10$\,ms). Subsequently,
  MHD- or viscosity-driven mass ejection occurs. The panel (a) shows a
  possible mass ejection history for the MNS formation case. Since
  both the MNS and surrounding disk are differentially rotating and
  strongly magnetized, MHD turbulence is likely to be generated. Then,
  the viscous effect in the MNS can be the cause of the early
  viscosity-driven mass ejection in $\alt 100$\,ms after
  merger. Subsequently, the viscous effect in the disk can drive mass
  ejection. Because of the presence of the MNS, which is a strong
  neutrino emitter, the neutrino irradiation plays a key role for
  determining the electron fraction of the ejecta. The panel (b) shows
  a possible mass ejection history for the prompt BH formation, for
  which only dynamical mass ejection and viscosity-driven mass
  ejection from the disk can occur, and the neutrino irradiation plays
  a minor role.
\label{fig3}}
\end{figure}

During and after neutron-star mergers, neutron-rich matter can be
ejected. First, at merger, the matter is dynamically ejected on the
timescale of $\alt 10$\,ms. Such mass ejection is referred to as the
dynamical mass ejection. Second, the mass ejection can proceed
from the merger remnant through MHD or viscous processes. Such mass
ejection is referred to as the post-merger mass ejection (see
Fig.~\ref{fig3} for these mass ejection processes).  In the following subsections,
we describe these two mass ejection mechanisms. We focus  on the mass, velocity, and electron fraction of the ejecta
because these quantities determine the property of EM counterparts
associated with the ejecta.

\subsection{Dynamical Mass Ejection from Binary Neutron Stars}\label{sec3.1}

In the mergers of binary neutron stars, strong shock waves are generated
by the high-velocity ($\sim 0.2c$) collision.
 In the shock waves, kinetic energy associated with the neutron
stars' plunging motion is converted to thermal energy, which enhances
thermal pressure and induces the ejection of the shocked matter. Also,
if an MNS is the merger remnant, it is initially highly
nonaxisymmetric and oscillating.  Such nonaxisymmetric MNS
gravitationally exerts torque on the matter surrounding it and induces
quick angular-momentum transport. Through this process, the matter in the
outer part of the system gains energy sufficient for ejection
from the system. These two mechanisms drive dynamical mass ejection.
The timescale of these processes is $\alt 10$\,ms.  Gravitational torque causes
matter to be ejected primarily in the equatorial direction, while 
shock heating causes it to be ejected in a less anisotropic manner.

\subsubsection{Mass}\label{sec3.1.1}

The mass of dynamical ejecta depends on the total mass, $m$, and mass
ratio, $q=m_2/m_1$, of binary neutron stars. 
For $m > M_{\rm thr}$, a BH is promptly formed after
the onset of merger (Fig.~\ref{fig1}). For $q \approx 1$, $\geq 99.9$\% of the
neutron-star matter is swallowed by the formed
BH~\cite{Shibata05,Kiuchi09}, and appreciable mass ejection cannot be
expected. If the mass ratio is different from unity, a fraction of
matter may be dynamically 
ejected~\cite{Hotoke13a,Dietrich17,Dietrich17b}. In this case, tidal
torque exerted by a deformed compact object collapsing to a BH is 
what  drives the dynamical mass ejection.  Numerical-relativity
simulations show that  $q\alt 0.8$ is necessary 
  for  dynamical mass ejection with mass $\geq
10^{-3}M_\odot$.

In the case of MNS formation, the dynamical ejecta mass depends strongly
on the neutron-star EOS as well as $m$ for the following reason:
For stiff EOSs (i.e., large neutron-star radii), the velocity of
two neutron stars at merger is relatively small because the minimum
orbital separation is large; thus, the shock heating efficiency
and oscillation kinetic energy of the remnant MNS are relatively
small. This results in a small dynamical ejecta mass. For practically
the same reason, the dynamical ejecta mass depends on the total mass
of the system, because for high total mass, the shock heating
efficiency and kinetic energy of the MNS oscillation can be large,
resulting in a high dynamical ejecta mass.  Numerical-relativity
simulations show that for EOSs with $R_{\rm NS} \agt 13$\,km or for
$m \alt 2.6M_\odot$, the dynamical ejecta
mass is of the order of $10^{-3}M_\odot$ for $q\sim 1$.  Only for $q
\alt 0.8$, the dynamical ejecta mass can be $\agt
0.005M_\odot$~\cite{Hotoke13a,Dietrich17,Dietrich17b}. By contrast,
 for $R_{\rm NS} \alt 12$\,km with $m \agt 2.7M_\odot$, the
dynamical ejecta mass could be $\sim 0.01M_\odot$ depending weakly on
$q$. Thus, the dynamical ejecta mass contains information about the
neutron star EOS.

\subsubsection{Velocity}\label{sec:velocity}

Since the dynamical mass ejection occurs from the vicinity of merged
objects of scale $R$, the velocity of the ejecta should be of the
order of its escape velocity, i.e., $\sim \sqrt{Gm/R} \approx
0.44c(m/2.6M_\odot)^{1/2} (R/20\,{\rm
  km})^{-1/2}$. Numerical-relativity simulations show that the typical
average velocity is 0.15--0.25$c$ for the case of  MNS formation (e.g.,
Ref.~\cite{Hotoke13a}). For the prompt formation of a BH from highly
asymmetric binaries, the average velocity of ejecta is higher $\sim 0.3c$,
because the dynamical mass ejection proceeds only for matter in the
vicinity of the object collapsing to a BH by a tidal torque exerted. 

In the case of the MNS formation, the dynamical mass ejection is induced in
part by shock heating. For the shocked ejecta component, a fraction of matter could
have a relativistic speed up to $\sim 0.8c$ and, in addition, the ejecta
morphology is quasi-spherical~\cite{Hotoke13a,Bauswein}.  Such
high-velocity ejecta can generate a characteristic observational
feature during the interaction with interstellar matter (ISM); cf. \S 4.3~\cite{Radice2018,Hotoke18}.

\subsubsection{Electron fraction}\label{sec:ele}

The electron fraction ($Y_e$) of ejecta is one of the key quantities
for determining the abundance of elements synthesized by r-process
nucleosynthesis~\cite{GBJ2011,Oleg,Wanajo14,LR15}. The abundance
pattern of the r-process elements is crucial for
determining the opacity of EM emission from the merger
ejecta~\cite{Metzger2010,BK2013,TH2013,Kasen15,Lippuner17}. 

Because the typical  $Y_e$ value for neutron stars is quite low,
0.05--0.1, $Y_e$ of the dynamical ejecta would  also be low if the
neutron-star matter is ejected without undergoing weak interaction
processes. However, the dynamical ejecta could be influenced strongly
by the weak processes.  First, shock heating at merger and during
subsequent evolution of the merger remnant increases the matter
temperature beyond 10\,MeV~\cite{Sekig2011,Sekig15}.  In such a
high-temperature environment, the electron-positron pair creation is
enhanced. As a result, neutrons easily capture positrons via $n + e^+
\rightarrow p + \bar\nu_e$.  Because the luminosities and average energies 
of electron neutrinos and electron antineutrinos are roughly equal, and
because the average energies are larger than the neutron-proton mass difference,
the $Y_e$ value of the initially neutron-rich material is driven toward $1/2$.
Thus, in the presence of many
positrons produced by  pair creation, the fraction of protons and
$Y_e$ are increased (i.e., the neutron-richness is reduced)~\cite{QW96}.

In the presence of an MNS that is a strong neutrino emitter, the
neutrino irradiation to the matter surrounding the MNS could
significantly change its composition. Since neutrons and protons
absorb neutrinos via $n + \nu_e \rightarrow p + e^-$ and $p +
\bar\nu_e \rightarrow n + e^+$, respectively, the fractions of
neutrons and protons tend to equilibrate. 
Because the luminosity and average energy of electron neutrinos and
electron antineutrinos from the MNS are not significantly different, the
fractions of protons and neutrons approximately approach the same
values (i.e., $Y_e$ approaches $1/2$ and the neutron-richness is 
significantly reduced).

As  mentioned above, there are two engines driving dynamical mass
ejection: shock heating and tidal torque.  
Both effects play an important role in  the case of MNS formation.  
On one hand, shock heating and
neutrino irradiation from the MNS     increase $Y_e$ for a large fraction of
ejecta. On the other hand, matter ejected
by tidal torque does not always undergo the weak interaction: If a
fraction of the matter is ejected by tidal torque without undergoing
shock heating and neutrino irradiation, the low-$Y_e$ state is
preserved. Therefore, the dynamical ejecta for the MNS formation case
in general has components with a wide range of $Y_e$ between $\sim
0.05$ (i.e., the original value in neutron stars) and $\sim 0.5$, and
if the weak-interaction effect is not significant, a large fraction of ejecta
has low values of $Y_e$.

In the case of prompt BH formation, most of the shock-heated matter is
swallowed by the BH, and a strong neutrino irradiation source such as 
an MNS is absent.  For asymmetric binaries, a fraction of matter is
ejected by the effect of tidal torque, but in this case, the weak
interaction does not play a role; therefore, $Y_e$ of the ejecta 
is low, $Y_e \alt 0.1$.

\subsection{Dynamical Mass Ejection from Black Hole-Neutron Star Binaries}\label{sec3.2}

If a neutron star is tidally disrupted by its companion BH, a fraction
of the neutron-star matter is ejected.  In contrast to binary neutron
star mergers, for BH-NS binaries, only the tidal effect plays an
important role in the dynamical mass ejection.

Broadly speaking, the mass of dynamical ejecta is determined by how
the tidal disruption of a neutron star proceeds. If a neutron star is
tidally disrupted far from the ISCO of its companion BH, a 
fraction of the neutron-star matter remains outside the BH
horizon after merger. Numerical-relativity simulations show that for such cases,
typically $\sim 20\%$ of the matter located outside the horizon
escapes from the system as ejecta~\cite{Kyutoku15}. Thus, larger disk
mass results in larger dynamical ejecta mass, up to $\sim 0.1M_\odot$
at maximum. Of course, if neutron stars are not tidally
disrupted, the dynamical ejecta mass is absent. Thus, the dynamical
ejecta mass is in the range of 0--$0.1M_\odot$ for BH-NS binaries.

The average velocity of dynamical ejecta is determined by the velocity
scale of the neutron star at tidal disruption, i.e.,
0.2--$0.3c$. Again, high-velocity matter can be present because a part
of ejecta comes from the vicinity of the BH horizon. In particular,
in the case of a spinning BH, the radius of the event horizon is small,
so the fraction of the high-velocity component can be
increased.

Because dynamical ejecta is launched predominately by tidal torque
and weak-interaction processes such as neutrino irradiation play  minor
roles in the ejecta, the $Y_e$ value of the dynamical ejecta is always low ($\alt 0.1$)
\cite{Foucart14,Foucart15,Kyutoku17}.  This
result is highly different from that in binary neutron star mergers 
resulting in an MNS (see \S~\ref{sec3.1}). 

\subsection{Viscosity-Driven Mass Ejection from Merger Remnants}\label{sec3.3}

In general, after merger of neutron-star binaries, an MNS or a BH surrounded by a
disk is  formed. At their formation, both the MNS and disk 
are differentially rotating and likely to be strongly magnetized; therefore,
MHD turbulence should be induced. Turbulent
viscosity could then be strongly enhanced as mentioned in
\S~\ref{sec2}. This viscous effect induces the so-called
viscosity-driven mass ejection~\cite{FM13,FM14,Just2015,Fujiba17}.  
We describe this  mechanism in the following subsections.

\subsubsection{Mass ejection driven by the viscous effect of massive neutron stars}\label{sec3.3.1}

First, we discuss the case of MNS formation  for binary neutron star
mergers.  If MHD turbulence develops and the resulting turbulent
viscosity is sufficiently high, the differential rotation energy of the remnant MNS could be the energy source of mass ejection. 
The angular momentum is transported in the MNS on the
timescale described by Eq.~\ref{tvis0}.  As a result, the angular velocity profile of the MNS is rearranged into  a
rigidly rotating state. The density and pressure
profiles also change during this transition  because the centrifugal force is
rearranged. Here, the total rotational kinetic energy of an MNS estimated by
$T_{\rm kin} \sim I\Omega^2/2 \sim 0.3M_{\rm
  MNS}R^2\Omega^2$~\cite{FIP86,CST94} is quite large:
\beq
T_{\rm kin} \sim 2 \times 10^{53}
\left({M_{\rm MNS} \over 2.6M_\odot}\right)
\left({R \over 15\,{\rm km}}\right)^2
\left({\Omega \over 7000\,{\rm rad/s}}\right)^2 \,{\rm erg}.
\eeq
This energy could be redistributed in the viscous timescale of
$\sim 10$--20\,ms.  In association with the change of the density profile,
strong density waves are generated. The density waves subsequently
propagate outward, and consequently, shocks are generated in the disk.  The
shock waves sweep matter into the disk and envelope, which subsequently
undergoes outgoing motion. If the energy of a fraction of the matter becomes
high enough, mass ejection could occur.

Because the power of the density waves depends on the strength of the viscous effect,
the ejecta mass in this process depends on the viscous
parameter. A numerical-relativity simulation shows that the eject mass
is $\sim 0.01M_\odot(\alpha_{\rm vis}/0.02)$~\cite{Fujiba17}.  The
ejecta is launched originally from the vicinity of the MNS.  Therefore,
the typical velocity of this ejecta component agrees approximately with the escape velocity of the MNS, i.e., $\sim
0.15c$. The electron fraction of this component is widely distributed as in 
dynamical ejecta.  However, the low-$Y_e$ components are
absent because the neutrino irradiation from the MNS is strong enough
to increase it to $Y_e \agt 0.2$, yielding values between $0.2$ and $0.5$.

\subsubsection{Mass ejection driven by the viscous effect of disks}\label{sec3.3.2}

For the longer-term evolution, viscous heating and angular momentum
transport in the disk play  important roles in mass ejection
regardless of the formation of MNS or BH.  In early
disk evolution, thermal energy generated by  viscous heating is
consumed primarily by neutrino emission. This stage is described
by a neutrino-dominated accretion disk~\cite{DiMatteo02} with a
fraction of the outflow toward the polar direction driven by neutrino
heating including neutrino-antineutrino pair annihilation heating.  In
the later stages, the mass, density, and temperature of the disk
decrease because of the outflow and accretion onto the MNS. The
decrease of the temperature, $T$, causes a reduction in 
the neutrino emissivity 
because of its strong dependence on $T$, which is approximately proportional to $T^6$~\cite{RJ96}.  Then,
the viscous heating is used primarily for the adiabatic expansion of
the disk toward the equatorial direction. The continuous viscous
heating causes the disk matter to eventually escape from the system as ejecta.


Because  viscous mass ejection from a disk should occur regardless
of the viscous parameter (for reasonably large values of $\alpha_{\rm
  vis}$), ejecta mass in this process depends weakly on its value.
Numerical simulations show that the ejecta mass could be a substantial
fraction (more than half) of the disk mass of 0.01--$0.1M_\odot$ for
the presence of an MNS~\cite{FM14,Perego,Fujiba17}.  For the presence
of a BH, the mass falling into the BH is larger than that of the
outflow. However, numerical simulations for disks around spinning BHs
show that $\sim 20\%$ of the disk mass can be ejected~\cite{FM13,Just2015}. 
If the matter is ejected efficiently
by MHD processes, this fraction may be increased by a factor of
two~\cite{SM17,Fetal18}.

The ejecta in this mechanism is launched primarily from the outer part
of disks. If the mass ejection occurs at a radius of $r \agt 100G
c^{-2}M$ ($M=M_{\rm MNS}$ or $M_{\rm BH}$), then the characteristic
velocity would be $\alt 0.1c$. Thus, the typical velocity of this
ejecta component is smaller than that for dynamical ejecta and early
viscosity-driven ejecta powered by the MNS.

The values of $Y_e$ within the dynamical ejecta vary widely. 
However, the low end depends strongly on the presence or
absence of the MNS, which can be the strong neutrino irradiation
source~\cite{FM14}. In the presence of the MNS, the low end of $Y_e$ could be 
$\sim 0.3$~\cite{Perego,Fujiba17}, whereas in its absence (i.e., in the
presence of a BH), low $Y_e$ values are preserved for a substantial
fraction of the ejecta~\cite{FM14,Just2015}. The reasons are 
that the disk is dense and electrons are degenerate, resulting in the
low $Y_e$ state in the disk, and that the weak interaction does not
play an important role because neutrino irradiation is weak in this
case~\cite{FM13}.

\subsection{Summary of Ejecta}\label{sec3.4}

Table 1 summarizes the typical properties of ejecta, 
showing that the ejecta quantities depend strongly on the binary
parameters, so that the observational features of the EM emission (in
particular kilonova emission: see \S\,\ref{sec4.2}) can be different
for each merger event. 

\begin{table}[bt]
\begin{center}
\caption{$M_{\rm ej,dyn}$ and $M_{\rm ej,vis}$: dynamical and
  post-merger ejecta mass in units of $M_\odot$, $Y_{e,{\rm dyn}}$:
  $Y_e$ of dynamical ejecta, $Y_{e, {\rm vis}}$: $Y_e$ of post-merger
  ejecta, $\langle v_{\rm ej} \rangle$: average velocity of dynamical
  ejecta in units of $c$. Low-$m$, Mid-$m$, and High-$m$ imply that
  the remnants soon after the merger are SMNS, HMNS, and BH.
  BNS denotes binary neutron star.
\label{tab1} }
\begin{tabular}{ccccccc}
Type of binary & Remnant & $M_{\rm ej,dyn}$ 
& $M_{\rm ej,vis}$ 
& $Y_{e,{\rm dyn}}$  & $Y_{e,{\rm vis}}$ 
& $\langle v_{\rm ej} \rangle$ \\
Low-$m$ BNS & SMNS & $O(10^{-3})$ & $O(10^{-2})$
& 0.05--0.5 & 0.3--0.5 & 0.15 \\
Mid-$m$ BNS (stiff EOS) 
& HMNS & $O(10^{-3})$ & $O(10^{-2})$
& 0.05--0.5 & 0.2--0.5 & 0.15 \\
Mid-$m$ BNS (soft EOS) 
& HMNS & $\sim 10^{-2}$ & $O(10^{-2})$
& 0.05--0.5 & 0.2--0.5 & 0.20 \\
High-$m$ BNS ($q \sim 1$)
& BH & $< 10^{-3}$ & $< 10^{-3}$
& --- & --- & --- \\
High-$m$ BNS ($q \ll 1$)
& BH & $O(10^{-3})$ & $\alt 10^{-2}$
& 0.05--0.1 & 0.05--0.3 & 0.30 \\
BH-NS
& BH & 0--$0.1$ & 0--$0.1$
& 0.05--0.1 & 0.05--0.3 & 0.30  
\end{tabular}
\end{center}
\end{table}

\section{ELECTROMAGNETC COUNTERPARTS OF NEUTRON-STAR MERGERS}\label{sec4}

Neutron star mergers eject a substantial amount of neutron-rich
material, in which r-process nucleosynthesis robustly occurs.
Subsequently, synthesized radioactive elements shine, in particular,
as a kilonova (macronova). In addition, the ejecta have large kinetic
energy with mildly-relativistic velocities, leading to a long-lasting
synchrotron remnant.  In the following subsections, we first summarize the general
properties of r-process nucleosynthesis in mergers, then describe
models of kilonovae and synchrotron remnants as promising EM signals.

\subsection{r-Process Nucleosynthesis and Ejecta Opacity}\label{sec4.1}

As described in \S~\ref{sec3}, dense neutron-rich matter is generally
ejected in neutron-star mergers. The neutron-rich ejecta can
subsequently synthesize heavy elements through r-process
nucleosynthesis,, that is, by rapid neutron capture, where the
capture timescale is typically shorter than the $\beta$-decay
timescales~\cite{QW96}.

In the r-process nucleosynthesis, the abundance of elements
synthesized depends primarily on the neutron richness, entropy, and
density~\cite{QW96}.  Among these properties, the neutron richness
(i.e., $Y_e$) is the key quantity in mergers.  Numerical calculations
show~\cite{GBJ2011,Oleg} that for ejecta only with neutron-rich matter
of $Y_e \alt 0.1$, r-process elements with mass number larger than $A
\agt 120$ (i.e., the elements in the so-called second and third peaks)
are robustly synthesized.  In this case, the mass fraction of
elements with $A \alt 120$ is quite small.  This finding implies that for
BH-NS mergers and binary neutron star mergers collapsing promptly to a
BH, predominantly heavy r-process elements are  synthesized.  By
contrast, from ejecta with $Y_e \agt 0.25$, heavy elements with $A
\agt 130$ (e.g., lanthanides) are not 
synthesized~\cite{Oleg,Wanajo14,LR15}.
In the presence of a wide range of $Y_e$  values in ejecta, r-process
elements with a wide mass range are synthesized, as was first pointed out
in Ref.~\cite{Wanajo14}. As mentioned in \S~\ref{sec3}, for mergers
of binary neutron stars leading to an MNS, matter with a wide range of
$Y_e$ values, 0.05--0.5 (see Table 1), is ejected; thus, r-process
elements with $A \agt 70$ are synthesized simultaneously. 

\subsection{Kilonova (Macronova)}\label{sec4.2}
Kilonovae are uv-optical-IR transients powered radioactively by
r-process elements freshly synthesized in merger ejecta.  

\subsubsection{Radioactive heating}\label{sec4.2.1}

The
radioactive decay channels of neutron-rich heavy elements are (i) $\beta$-decay, (ii) $\alpha$-decay, and (iii) fission. The specific heating rate
of the second and third of these channels depends sensitively on the abundance of
superheavy nuclei ($A\geq 210$ for alpha decay and $A\geq250$ for
fission). In the following, we describe the heating
process for each channel.


{\em $\beta$-decay:} $\beta$-unstable nuclei decay toward the stability valley
without changing the atomic mass number.  Because the number of nuclei
is conserved for each atomic mass number, the decay rate is approximately
proportional to $t^{-1}$, where radioactive species with mean-lives
$\tau\sim t$ predominately contribute to the decay rate at $t$.  The
electron energy liberated in each decay generally decreases with the
lifetime as $E_e \propto \tau^{-1/5}$ to $\tau^{-1/3}$. Thus, the
energy releasing rate in $\beta$-decay electrons per unit mass is
$\dot{q}_e(t)\propto t^{-6/5}$ to
$t^{-4/3}$~\cite{Metzger2010,Hotoke17}, which is typically written as
\begin{eqnarray}
\dot{q}_e(t) \approx 3 \cdot 10^{9}\,{\rm erg\,s^{-1} g^{-1}}\,
\left(\frac{t}{\rm 1\,day}
\right)^{-4/3}.
\end{eqnarray}
$\beta$-decay is often followed by $\gamma$-ray emission and the
efficiency of the energy release in $\gamma$-rays is $\sim 0.3$--$2$
times that of $\dot{q}_e(t)$. 

\begin{figure*}[t]
\includegraphics[width=120mm]{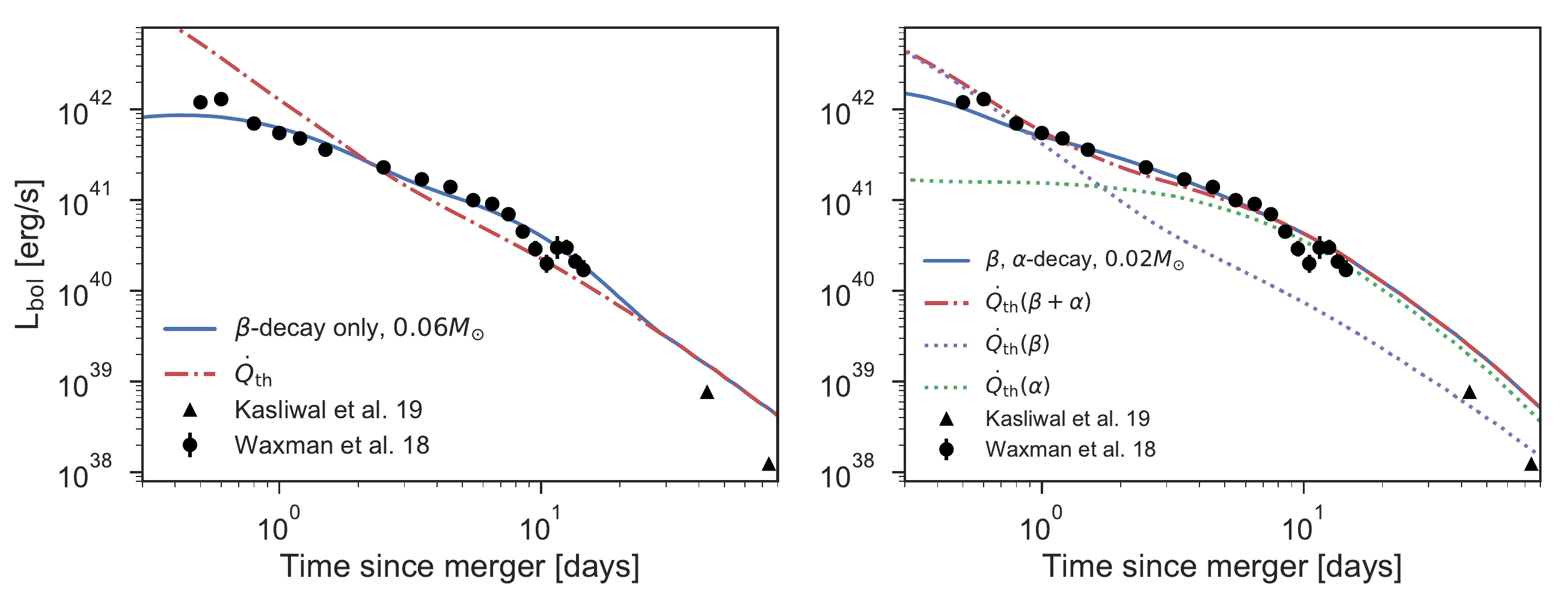}
\caption{$\beta$-decay heating rate and bolometric light curve models for
  ejecta of the solar r-process abundance with $A\geq 85$ ({\it left})
  and those including $\alpha$-decay heating ({\it right}).  For the {\it
    left} panel, the total r-process mass and the typical ejecta
  velocity are set to be $0.06M_{\odot}$ and $0.15c$. The opacity is
  assumed to be $1\,{\rm cm^2/g}$ for $v>0.15c$ and $7\,{\rm cm^2/g}$
  for $v\leq 0.15c$. Note that this opacity distribution
  is  phenomenological to fit the bolometric data and somewhat motivated 
  from the models shown in Refs. \cite{EM10,Nakar2018}. 
  For the {\it right} panel, the initial abundance
  of $A=222$, $223$, $224$, and $225$ is taken to be $Y_A=4.0\times
  10^{-5}$, $2.7\times 10^{-5}$, $4.1\times 10^{-5}$, and $2.7\times
  10^{-5}$, respectively, corresponding to the DZ31 model presented in
  Wu et al. \cite{Wu18}.  The total r-process mass and the typical
  ejecta velocity are supposed to be $0.02M_{\odot}$ and $0.1c$ for
  this case. The opacity is assumed to be $0.1\,{\rm cm^2/g}$ for
  $v>0.1c$ and $1\,{\rm cm^2/g}$ for $v\leq 0.1c$.  Also depicted are
  the observed bolometric light curve data of
  GW170817~\cite{Waxman2018} and $\nu L_{\nu}$ of the late-time
  Spitzer observations at $4.5\,{\rm \mu m}$ \cite{Kasliwal18}. 
\label{fig_bol}}
\end{figure*}

{\em $\alpha$-decay:} Neutron-rich elements with $210 \leq A\lesssim 254$
increase their proton fraction through $\beta$-decay until the point at
which they are predominately disintegrated by $\alpha$-decay. After a
number of $\alpha$-decays and $\beta$-decays, they eventually reach stable nuclei with
$A<210$.  Each $\alpha$ decay liberates energy of $\sim 5$--$10$ MeV.
Among the $\alpha$-unstable elements, $^{222}$Rn, $^{223}$Ra, $^{224}$Ra,
$^{225}$Ra, and $^{225}$Ac are particularly relevant for the kilonova
heating rate \cite{Wu18}.  
 In a decay chain of these elements, $20$--$30$ MeV is
released in total.  Nuclei with
$222\leq A\leq 225$ can dominate over the $\beta$-decay heating for $t>$
a few days, if the total mass of these elements is 
$\gtrsim 10^{-3}M_{\odot}$ (Fig.~\ref{fig_bol}). 

{\em Spontaneous fission:} Transuranium nuclides with $A\gtrsim 250$
may be disintegrated by spontaneous fission, in which $\sim
100$--$200$\,MeV is released as kinetic energy of fission fragments.
Thus, the energy release of each fission is greater by a factor of
$\gtrsim 100$ than in $\beta$-decay.  Although the $Q$-value and half-life
of spontaneous fission, as well as the abundance of transuranium
nuclides synthesized in merger ejecta, are highly uncertain,
spontaneous fission could potentially dominate the heating rate at later
times of $\gtrsim 10$\,day~\cite{Wanajo14,Hotoke16b,Zhu18,Wu18}.  For
instance, a notable element is $^{254}$Cf, of which the
half-life is $60.5$\,day and the $Q$-value is $185$ MeV
\cite{Zhu18,Wu18}.

High energy charged particles (electrons, $\alpha$-particles, and fission
fragments) produced by radioactive decay deposit their kinetic energy
to the thermal energy of merger ejecta on the following
timescale~\cite{Barnes16,Kasen18},
\begin{eqnarray}
t_{\rm th} \approx \left(\frac{\sigma_{\rm st}(E_i)N v_i}{E_i} \right)^{-1},
\end{eqnarray}
where $\sigma_{\rm st}$ is the stopping power determined predominantly 
by the collisional ionization and excitation of ions, $N$ is the
number density of ions, $E_i$ and $v_i$ are the initial kinetic energy and
velocity of a particle, respectively. $E_i$ is typically $0.1$--$1$
MeV for electrons, $5$ MeV for $\alpha$-particles, and $100$ MeV for
fission fragments. Since the density decreases with time in expanding
ejecta, the thermalization time increases and eventually exceeds the
expansion time.

The thermalization timescales for beta-, alpha-, and gamma--decay are 
\begin{eqnarray}
t_{\rm th,\beta} & \sim & 30\,{\rm day}\,
\left(\frac{M_{\rm ej}}{0.05M_{\odot}} \right)^{1/2}
\left(\frac{v_{\rm ej}}{0.1c} \right)^{-3/2} 
\left(\frac{E_{i}}{0.5\,{\rm MeV}} \right)^{-1/2},\\
t_{\rm th,\alpha} & \sim & 45\,{\rm day}\,
\left(\frac{M_{\rm ej}}{0.05M_{\odot}} \right)^{1/2}
\left(\frac{v_{\rm ej}}{0.1c} \right)^{-3/2} 
\left(\frac{E_{i}}{5\,{\rm MeV}} \right)^{-1/2}, \\
t_{\rm th,\gamma} 
&\sim & 2.4\,{\rm day}\,
\left(\frac{\kappa_{\gamma}}{0.05\,{\rm cm^2/g}} \right)^{-1}
\left(\frac{M_{\rm ej}}{0.05M_{\odot}} \right)^{1/2}
\left(\frac{v_{\rm ej}}{0.1c} \right)^{-1},
\end{eqnarray}
where $M_{\rm ej}$ and $v_{\rm ej}$ 
denote the mass and typical velocity of the ejecta, 
and $\kappa_{\gamma}\approx 0.05\,{\rm cm^2/g}$ is the mass
absorption coefficient of r-process elements at $\gamma$-ray energy of
$\sim 1$\,MeV. Note that the thermalization time for fission fragments, $t_{\rm th,sf} $, is $\sim 2t_{\rm th,\alpha}$ \cite{Barnes16}.

Once $t>t_{{\rm th},a}$~($a=\alpha$, $\beta$, $\gamma$, sf) is achieved, the
thermalization rate becomes lower than the adiabatic cooling rate; therefore,
a significant fraction of the radioactive energy is lost
adiabatically for charged particles and the $\gamma$-ray heating rate declines exponentially.  
The heating rate is equal to the 
energy generation rate for $t\ll t_{{\rm th},a}$, while, for $t\gtrsim t_{{\rm th},a}$, the heating rate
 deviates from  the energy generation rate \cite{Waxman2018,Kasen18} and goes approximately as $\propto t^{-3}$ 
 for $t\gg t_{{\rm th},a}$ \cite{Waxman2018}.






\subsubsection{Opacity}\label{sec4.2.2}

The opacity for photons plays an essential role for the light curves
and spectra of kilonovae. In kilonovae, the opacity is determined
primarily by the bound-bound absorption of heavy
elements~\cite{opacity,TH2013}.  Notably, the
bound-bound absorption opacity of open $f$-shell elements (lanthanides
and actinides) differs significantly  from the opacity of others,
because  open $f$-shell elements have such a high number of
 excited levels 
with relatively low excitation energy
that  the number of transition lines in the optical and IR
bands is greatly enhanced~\cite{opacity,BK2013,TH2013}.  Radiation transfer simulations of
merger ejecta show that the mean opacity, $\kappa$, is $\gtrsim
10\,{\rm cm^2/g}$ for lanthanide-rich ejecta while it is $\sim
0.1\,{\rm cm^2/g}$ for lanthanide-free
ejecta~\cite{opacity,BK2013,TH2013,Kasen15,T2018,Wollaeger2018}.  This finding
implies that the $Y_e$ distribution of ejecta, which primarily
determines the abundance pattern of r-process elements, is the key for
determining the features of kilonovae.

\subsubsection{Typical light curve}\label{sec4.2.3}

For merger ejecta, the light curve peaks on a timescale \cite{Metzger2010}
\begin{eqnarray}
t_{\rm p} \approx \sqrt{\frac{\kappa M_{\rm ej}}{4\pi c v_{\rm ej}}}
\approx 10\,{\rm day}\, \left(\frac{\kappa}{10{\rm \,cm^2/g}}\right)^{1/2}
\left(\frac{M_{\rm ej}}{0.04M_{\odot}}\right)^{1/2}
\left(\frac{v_{\rm ej}}{0.1c}\right)^{-1/2}.\label{eq:tp}
\end{eqnarray}
The luminosity and effective temperature are estimated as
\begin{eqnarray}
L_{\rm bol} (t_{\rm p}) & \approx & \dot{Q}_{\rm th}(t_p) 
= M_{\rm ej}\cdot \dot{q}_{\rm th}(t_p)
\approx 4\cdot 10^{40}\,{\rm erg/s}\,
\left(\frac{t_p}{10\,{\rm day}} \right)^{-1.3}
\left(\frac{M_{\rm ej}}{0.04M_{\odot}}\right),\\
T_{\rm eff} (t_{\rm p}) & \approx &  
\left(\frac{L_{\rm bol}(t_{\rm p})}{4\pi \sigma v_{\rm ej}^2 t_{\rm p}^2} 
\right)^{1/4} \approx 2000\,{\rm K}\, 
\left(\frac{L_{\rm bol,p}}{4\cdot 10^{40}\,{\rm erg/s}} \right)^{1/4}
\left(\frac{v_{\rm ej}}{0.1c} \right)^{-1/2}
\left(\frac{t_{\rm p}}{10\,{\rm day}} \right)^{-1/2},
\end{eqnarray}  
where $\sigma$ is the Stefan-Boltzmann constant. These equations show
that lanthanide-free ejecta are brighter, bluer, and peaks earlier than
lanthanide-rich ejecta  if the mass, velocity, and specific heating
rate are the same.

Figure~\ref{fig_bol} ({\it left}) shows the $\beta$-decay heating rate
for the solar r-process abundance pattern with $A\geq 85$ and a bolometric
light curve calculated using a simple one dimensional ejecta model, in
which $\kappa$ is assumed to be $1\,{\rm cm^2/g}$ for $v>0.15c$ and
$7\,{\rm cm^2/g}$ for $v\leq 0.15c$ with $M_{\rm ej}=0.06M_{\odot}$.
Note that this opacity distribution is 
phenomenological to fit the bolometric data and is somewhat motivated 
  from the models presented in, e.g., Refs. \cite{EM10,Nakar2018}. 
Kilonova bolometric light curves have following generic features.  The
bolometric luminosity is lower than the heating rate in the early
phase in which most part of the ejecta is optically thick.  
When the optical depth becomes below $\approx c/v_{\rm ej}$,
 photons in the entire ejecta start diffusing out from the ejecta 
 without significant adiabatic losses. 
At later times, the
ejecta density becomes so low that most of photons in the ejecta
diffuse out within one dynamical time thereby the
bolometric luminosity approaches  approximately the total heating rate.

Figure~\ref{fig_bol} ({\it right}) illustrates the case in which $\alpha$-decay
 enhances the kilonova heating rate.  In the example shown in the figure, the
$\alpha$-decay heating rate of the DZ31 model shown in Ref. \cite{Wu18}
is added to the $\beta$-decay heating rate.  Note that this model
predicts the production of much larger amounts of $\alpha$-unstable
nuclei than other nuclear mass models~\cite{Wu18}. With this
model, the ejecta mass of $\approx 0.02M_{\odot}$ is sufficient to
generate bolometric light curve as bright as the light curve with
only $\beta$-decay and $M_{\rm ej} \approx 0.06M_{\odot}$.


\subsection{Synchrotron Emission}\label{sec4.3}

The interaction of merger ejecta with the surrounding ISM
 produces a long-lasting synchrotron emission observable
in multi-wavelength bands from radio to X-rays~\cite{NP2011}. Various
types of merger ejecta, including dynamical and post-merger ejecta, sGRB
jets, and cocoons, can produce such signals. Here, we focus on the signal
arising from the dynamical ejecta because it is
closely related to the merger dynamics~\cite{Hotoke18,Hotoke16r}.

We  estimate  the flux from dynamical ejecta by modeling it as a
spherical expanding shell with single velocity and neglecting
relativistic corrections.  An ejecta with kinetic energy, $E$, and
initial velocity in units of $c$, $\beta_i$, expanding in the
surrounding ISM of a constant number density, $n$, is decelerated on the following
timescale:
\begin{eqnarray}
t_{\rm dec} \approx 30\,{\rm day}\,
\left(\frac{E}{10^{49}\,{\rm erg}} \right)^{1/3}
\left(\frac{n}{1\,{\rm cm^{-3} }} \right)^{-1/3}\beta_i^{-5/3}.
\end{eqnarray}
The ejecta velocity (in units of $c$), $\beta$, is constant for
$t<t_{\rm dec}$ and decreases as $\propto t^{-3/5}$ for $t\gtrsim t_{\rm
  dec}$ during the adiabatic expansion phase.  The light curve has a
peak at $t\sim t_{\rm dec}$\footnote{The peak time is longer than
  $t_{\rm dec}$ when the synchrotron self-absorption is
  important. Such a delay of the peak can occur for mergers at high
  ISM densities ($\gtrsim 1\,{\rm cm^{-3}}$) and/or low observed
  frequencies ($\lesssim 1$ GHz).}:
\begin{eqnarray}
F_{\nu,{\rm peak}} & \approx & 3\,{\rm mJy}\,
\left(\frac{E}{10^{49}\,{\rm erg}} \right)
\left(\frac{n}{1\,{\rm cm^{-3} }} \right)^{(p+1)/4}
\left(\frac{\epsilon_B}{0.1 } \right)^{(p+1)/4}
\left(\frac{\epsilon_{e}}{0.1}\right)^{p-1}
\beta_i^{(5p-7)/2}\nonumber \\
& & ~~~~~~~~~~~~~~~~\times \left(\frac{D}{100\,{\rm Mpc}}\right)^{-2}
\left(\frac{\nu}{1.4\,{\rm GHz}}\right)^{-(p-1)/2} , 
\end{eqnarray}
where $\epsilon_{B}$ and $\epsilon_{e}$ are the conversion
efficiencies of internal energy of the shocked ISM to magnetic-field
energy and accelerated electron energy, respectively, and $p$ is the
power-law index for the distribution function of accelerated
electrons.  The value of $p$ is likely to be $2$--$3$ as inferred from
the GRB afterglow and radio-supernova observations. Notably, the peak flux is quite sensitive to the ejecta velocity. For
a given ejecta mass, the flux increases with velocity as $\propto
\beta^{4.75}$ for $p=2.5$; therefore, the detecting such signals would
prove the velocity profile of merger ejecta.


The above estimate is valid if the following three conditions are
satisfied: (i) The self-absorption is negligible ($\nu>\nu_a$), (ii)
the observed frequency is above the characteristic synchrotron
frequency, ($\nu>\nu_m$), and (iii) the observed frequency is below
the synchrotron cooling frequency, ($\nu<\nu_c$). Here, the
characteristic synchrotron frequency 
and the cooling frequency are given, respectively, by
\begin{eqnarray}
&&\nu_m \approx 1\,{\rm GHz}\,\left(\frac{n}{1\,{\rm cm^{-3} }} \right)^{1/2}
\left(\frac{\epsilon_B}{0.1 } \right)^{1/2}
\left(\frac{\epsilon_{e}}{0.1}\right)^2\beta^5, \\
&&\nu_c \approx 10^{14}\,{\rm Hz}\,
\left(\frac{n}{1\,{\rm cm^{-3} }} \right)^{-3/2}
\left(\frac{\epsilon_B}{0.1 } \right)^{-3/2}
\left(\frac{t}{30\,{\rm d}}\right)^{-2}\beta^{-3},
\end{eqnarray}
and the self-absorption frequency at $t_{\rm dec}$ is estimated by
\begin{eqnarray}
\nu_{a,{\rm dec}} \approx 1\,{\rm GHz}
\left(\frac{E}{10^{49}\,{\rm erg}} \right)^{\frac{2}{3(p+4)}}
\left(\frac{n}{1\,{\rm cm^{-3} }} \right)^{\frac{3p+14}{6(p+4)}}
\left(\frac{\epsilon_B}{0.1 } \right)^{\frac{p+2}{2(p+4)}}
\left(\frac{\epsilon_{e}}{0.1}\right)^{\frac{2(p-1)}{p+4}}
\beta_0^{\frac{15p-10}{3(p+4)}}.
\end{eqnarray}
The above equations show that
$\nu_m$ and $\nu_a$ are typically lower than the radio frequency for
sub-relativistic ejecta with $n\lesssim 1\,{\rm cm^{-3}}$ and
that the cooling break is expected to occur between the optical and
X-ray bands. 

\begin{figure*}[t]
\includegraphics[width=120mm]{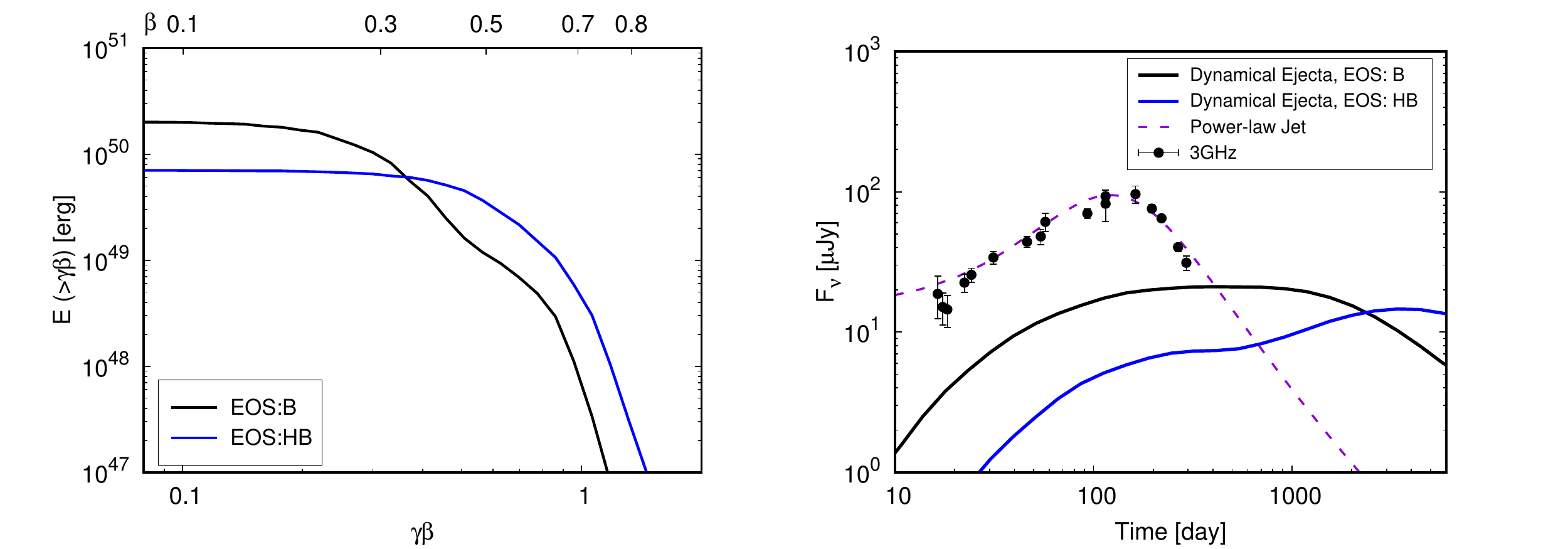}
\caption{Kinetic energy of dynamical ejecta as a function of four
  velocity ({\it left}) and afterglow light curves ({\it right}). Here
  the results of dynamical ejecta from binary neutron star mergers
  with two different EOSs (B and HB) and with mass
  $m_1=m_2=1.35M_\odot$ are shown. For computing these light curves,
  we employ $n=10^{-3}\,{\rm cm^{-3}}$ and microphysics parameters of
  $\epsilon_B=\epsilon_e=0.1$ and $p=2.2$ \cite{Hotoke18}. Also
  depicted are the observed data of the afterglow in GW170817 at $3$
  GHz and a light curve for a power-law structured jet model which
  agrees with the light curve data~\cite{Mooley18c} and
  the observed superluminal motion~\cite{Mooley18b}.
\label{fig_radio}}
\end{figure*}

As discussed in~\S\ref{sec:velocity}, the velocity of dynamical ejecta
is typically $\sim 0.2c$ and can reach up to $\sim 0.8c$. The
total kinetic energy is $\sim 10^{50}$--$10^{51}$\,erg and kinetic
energy in the fast component with $v\gtrsim 0.7c$ is $\sim
10^{47}$--$10^{49}$\,erg depending on each mass of the binary and
neutron-star EOS (see Fig.~\ref{fig_radio} and Ref. \cite{Radice2018,Hotoke18}). Such a velocity
distribution results in a relatively flat and years-lasting afterglow
light curve.

\section{GW170817}\label{sec5}

The observations of EM counterparts to GW170817
have, for the first time,  provided valuable
information  to test theoretical predictions for 
mass ejection and associated EM emission. In this section, we
summarize the observational features of the EM counterparts and
briefly describe theoretical models that are broadly consistent with
the observational results.

\subsection{Kilonova Observation}\label{sec5.1}

\begin{figure}[t]
\begin{center}
\vspace*{-15mm}
\includegraphics[width=120mm]{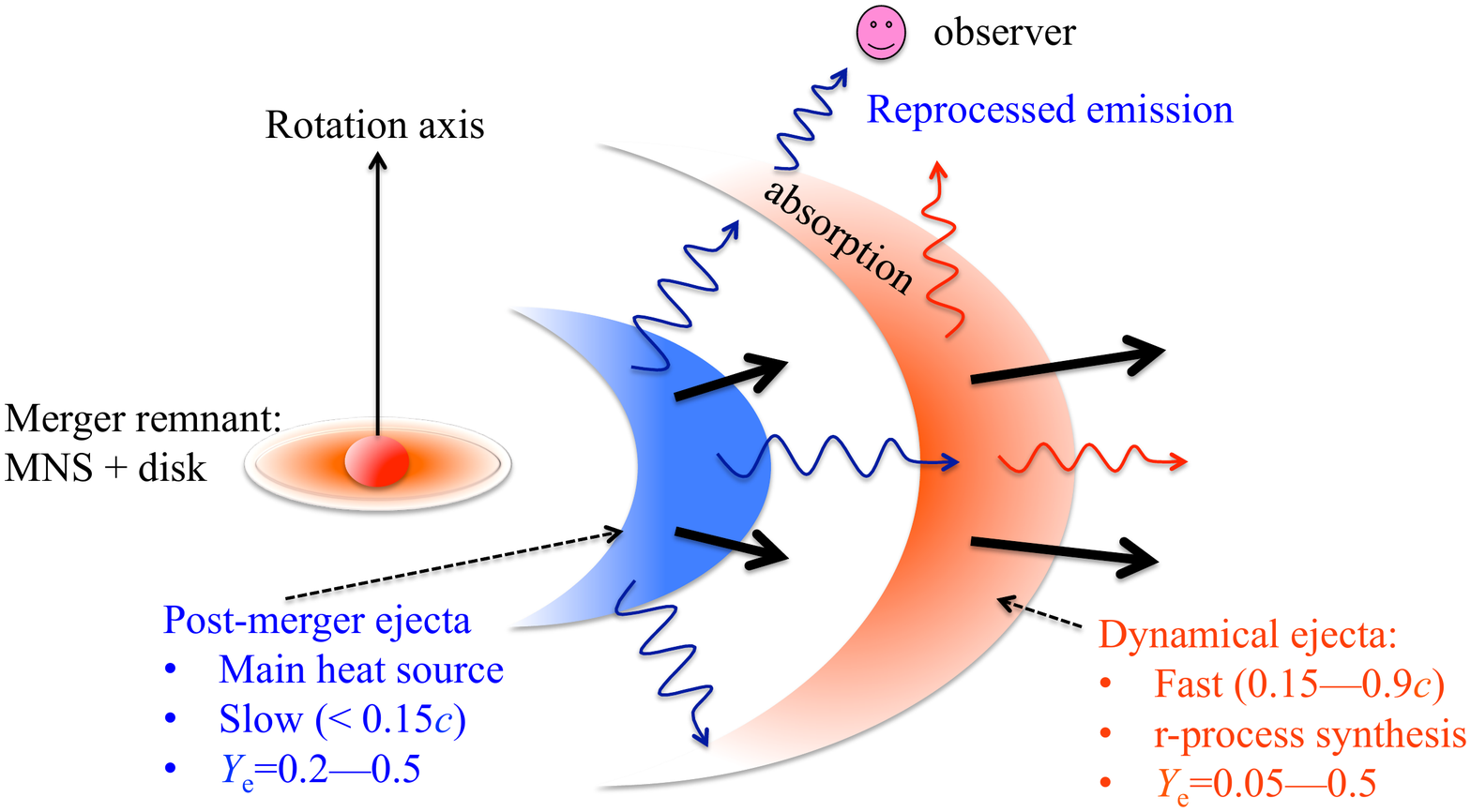}
\vspace*{-13mm}
\caption{Schematic picture of the ejecta profile for the case that a
  long-lived MNS is formed as a remnant. The outer falcate component
  denotes the neutron-rich dynamical ejecta. The inner falcate
  component denotes the less neutron-rich post-merger ejecta which is
  slower than the dynamical ejecta. Note that the gravitational-wave 
observation indicates that the merger remnant of GW170817 is observed 
along the direction of $\theta \alt 30^\circ$ from the rotation axis.}
\label{fig6}
\end{center}
\end{figure}

Figure~\ref{fig_bol} shows
the observed bolometric light curve data of
GW170817~\cite{Waxman2018,EM4,EM6,EM8, Arcavi2018} and $\nu L_{\nu}$
of the late-time Spitzer observations at $4.5\,{\rm \mu
  m}$~\cite{Kasliwal18}. Here, the
late-time Spitzer data are considered approximately as  the
bolometric luminosity. The observed data are largely consistent with
the $\beta$-decay heating with $M_{\rm ej}=0.06M_{\odot}$. Two notable
 features of this kilonova is that the light curve peaks at
$\lesssim 0.5$\,d and that the peak luminosity reaches $\sim 10^{42}$ erg/s.
As shown by Eq.~\ref{eq:tp}., this fact requires that some fractions
of the ejecta have a low opacity $\lesssim 1\,{\rm cm^2/g}$,
suggesting that there exists a substantial amount of material with a
quite low or even zero lanthanide fraction.  By contrast, the
evolution of the temperature (spectrum) at later times indicates the
existence of a lanthanide-rich component.  Therefore, the kilonova in
GW170817 shows us evidence that merger ejecta has components
with a broad range of $Y_e$ (e.g.,~Fig.~\ref{fig6}).


\begin{figure}[t]
\begin{center}
\includegraphics[width=90mm]{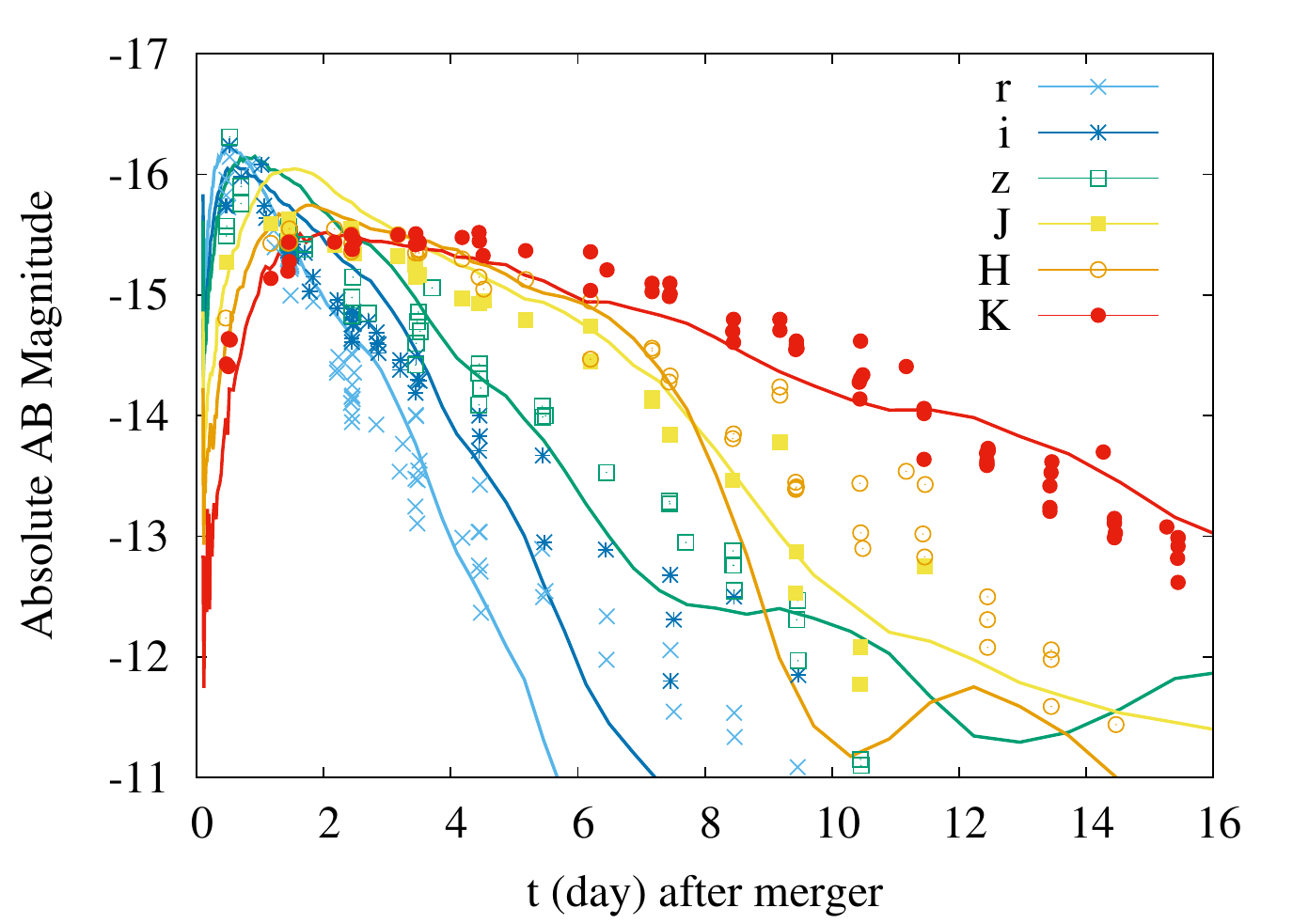}
\caption{Optical and near infrared (r, i, z, J, H, K bands) light
  curves of the kilonova associated with GW170817 (points) and
  theoretical model light curves (curves) based on
  numerical-relativity simulations (see Fig.~\ref{fig6}) assuming the
  viewing angle of $\approx 25^\circ$~\cite{Kawaguchi2018}.  The
  optical and near infrared data points are taken from
  Ref.~\cite{EM10}.  All of the magnitudes are given in AB
  magnitudes. }
\label{fig7}
\end{center}
\end{figure}
 
As discussed in~\S\,\ref{sec:ele}, dynamical ejecta has a wide range
of $Y_e$ values, but numerical-relativity simulations show that dynamical
ejecta mass would be $\alt 10^{-2}M_{\odot}$, which is smaller  
by a factor of $2$ or more
than that required to reproduce the observed luminosity. This
suggests that the merger remnant would eject $\gtrsim 0.03M_{\odot}$
from the remnant MNS and/or accretion disk (see \S\,3).  
The origin of different $Y_e$ components is and how they
are spatially distributed are under debate. Several models suggested
to date are as follows. 

 {\it Angular structure model}: a lanthanide-free component (blue) and
  a lanthanide-rich one (red) are angularly separated, for instance, the polar
  ejecta is lanthanide free~\cite{Kasen2017}. Fitting the photometric
  light curve data of GW170817 leads to the mass of $\approx
  0.01M_{\odot}$ and velocity of $0.3c$ for the blue component and the
  mass of $\approx 0.04M_{\odot}$ and velocity of $0.1c$ for the red
  component~\cite{EM8}. Introducing another component results in a
  better fit to the data~\cite{EM10}. 

 {\it Radial structure mode}: the composition (opacity) varies with the
  ejecta velocity, e.g., the opacity of the fast (slow) moving
  material is $0.8\,(5)\,{\rm cm^2/g}$, where the two components are
  separated at $v=0.1c$ \cite{EM6,Nakar2018}. 

{\it Temporal variation model}: the opacity evolves with time, which
  is expected from the time variation of the temperature and density
  of the ejecta~\cite{Waxman2018}.  The form
  $\kappa=\kappa_M(t/t_M)^{\gamma}$ is applied to GW170817 and
  $\kappa_M\approx 0.3\,{\rm cm^2/g}$, $\gamma \approx 0.6$, and
  $t_M\approx 1$~d. 

{\it Model motivated by numerical relativity}: this model employs two
  (or three) ejecta components motivated by the results of
  numerical-relativity simulations for the merger and
  post-merger~\cite{Perego2017,Shibata2017,Kawaguchi2018}. On the basis of
  the numerical results, the composition is varied both
  radially and angularly, and non-trivial radiation transfer
  effects are taken into account. 

Figure~\ref{fig7} compares optical and near-IR light curves of
the kilonova associated with GW170817 and theoretical curves derived
by a radiation-transfer simulation in the background of an ejecta
model obtained from numerical-relativity simulations (see
Fig.~\ref{fig6} for a schematic figure). This figure illustrates that
this model works
well~\cite{Kawaguchi2018}. However, it is not yet clear whether every
kilonova agrees with the prediction of numerical relativity, and
comparison with a number of future events is clearly needed to
establish the standard picture for kilonovae.

Before closing this section, we note that $\alpha$-decay and spontaneous
fission can potentially enhance the heating rate at late times (see
Fig.~\ref{fig_bol} for $\alpha$-decay).  Although we cannot conclude
whether or not such heavy elements play a role for the EM emission 
of GW170817, the estimated ejecta mass is significantly reduced from
$\approx 0.05M_{\odot}$ if these decay channels are important.  In 
future events, it may be possible to identify a signature of heavy
elements using a bolometric light curve at late times $\gg 10$\,day.

\subsection{Synchrotron Emission and Jet}\label{sec5.2}

The X-ray and radio afterglows of GW170817 were discovered at 9 and
16\,d after the merger \cite{Troja17,Hallinan17}.  The light curves
rise as $\propto t^{0.8}$ until $\approx 150$\,days~\cite{Mooley18a} and
then both X-ray and radio light curves  fall quickly as
$\propto t^{-2.2}$ \cite{Troja18,Mooley18c}. The spectrum of the
afterglow is consistent with a single power law, $F_{\nu}\propto
\nu^{-0.6}$, from the radio to X-ray bands \cite{Margutti18}, which is
described well by synchrotron radiation emitted by accelerated
electrons in the shocked ISM.  The slow rise over a timescale of
$150$~days is attributed to the fact that the jet structure includes a cocoon component
 and this feature is quite different from the typical GRB
afterglow light curve. It is also remarkable that the fast decline of
the light curve agrees with the light curve predicted for the
post-jet break regime of collimated jet models. Furthermore, the Very Long Baseline Interferometry
observations reveal that the unresolved radio emitting region exhibits
a superluminal motion with a Lorentz factor of $\approx
4$~\cite{Mooley18b}. These observational features confirm that the
afterglow arises from a narrowly collimated relativistic jet with some
structure seen from off-axis.  The kinetic energy and jet-half opening
angle are estimated as $E_j \approx 10^{49}$--$10^{50}$ erg and
$\theta_j \lesssim 5^{\circ}$, respectively \cite{Mooley18b}. Figure~\ref{fig_radio}
shows the light curve of a power-law structured jet model with
$E_j\approx 2\times 10^{49}$\,erg, $\theta_j \approx 3^{\circ}$,
$n\approx 10^{-3}\,{\rm cm^{-3}}$, and the viewing angle $\approx
21^{\circ}$.

Another important observation that is likely related to the jet is GRB
170817A detected at $1.7$\,s after the merger~\cite{GRB170817A},
which is much weaker than the typical sGRB.  This prompt $\gamma$-ray
emission requires a relativistic motion of the emission region \cite{Gottlieb18}. The
delay of the $\gamma$-ray detection from the merger indicates that the
jet should be formed for $\ll 1.7$ s after the merger.  
On the basis  of numerical-relativity simulations,
it has been suggested
 that a relativistic jet may be
driven by magnetic fields after an MNS collapses to a
BH~\cite{Ruiz2017} in a lifetime of $\ll 1.7$\,s. The collimation of
the jet in GW170817 can be interpreted as follows.  The jet interacts
with the material ejected around the polar region before the jet breaks out
from the ejecta surface. Consequently, the ejecta shocked by the jet
form a cocoon which helps collimation of the
jet~\cite{Nagakura2014,M2014,Duffell2015}. The small opening angle of
the jet in GW170817 indicates that an appreciable amount of ejecta is
present around the polar region prior to the jet formation\cite{Gottlieb18}.

Figure~\ref{fig_radio} depicts models for the afterglow light curves arising from dynamical ejecta
with an ISM of density $n=10^{-3}\,{\rm cm^{-3}}$.
 If the microphysics parameters are somewhat
optimistic, the radio emission with a flux density of $\approx
10\,{\mu}$Jy may be detectable in near future. Also, we
note that for future merger events, this radio emission may be a
primary target for the radio-band observation, if the viewing angle of
the merger events is sufficiently wide. 


\section*{ACKNOWLEDGMENTS} 
We thank  S. Fujibayashi, G. Hallinan, K.  Ioka, M. M. Kasliwal, K. Kawaguchi, K. Kiuchi, K. Kyutoku,
 K. P. Mooley, E. Nakar, T. Piran, D. Radice, Y. Sekiguchi, M. Tanaka, and S. Wanajo for useful discussions. 
This work was supported by Grant-in-Aid for Scientific Research 
(16H02183) of JSPS/Japanese MEXT. K.H. is supported by Lyman Spitzer Jr. Fellowship
at the Department of Astrophysical Science, Princeton University.


\end{document}